\documentclass[prl,superscriptaddress,showpacs,twocolumn]{revtex4}

\usepackage{amsmath,amssymb,amsfonts}
\usepackage{graphicx}
\usepackage{mathrsfs}
\usepackage{dcolumn}

\def\beq{\begin{equation}}
\def\eeq{\end{equation}}
\def\a{\alpha}
\def\b{\beta}
\def\g{\gamma}
\def\o{\omega}
\newcommand\psida{\psi^\dagger_{\alpha}}
\newcommand\psidb{\psi^\dagger_{\beta}}

\newcommand\psib{\psi_{\beta}}
\newcommand{\bk}{\mathbf{k}}
\newcommand{\bq}{\mathbf{q}}
\newcommand{\br}{\mathbf{r}}
\newcommand{\bR}{\mathbf{R}}
\newcommand{\Dk}{\mathbf{q}}
\newcommand{\Dom}{\Omega}
\newcommand{\DomT}{\Omega_T}

\newcommand{\xik}{\xi_{\mathbf{k}}}
\newcommand{\kF}{\mathbf{k}_F}
\newcommand{\vF}{\mathbf{v}_F}
\newcommand{\Gk}{\Gamma_{\mathbf{k}}}
\newcommand{\zk}{Z_\bk}

\newcommand{\ekb}{\varepsilon_{\mathbf{k}\beta}}
\newcommand{\ekab}{\varepsilon_{\mathbf{k}\alpha,\beta}}
\newcommand{\eb}{\varepsilon_{\beta}^0}
\newcommand{\ea}{\varepsilon_{\alpha}^0}
\newcommand{\omt}{\omega_0}
\newcommand{\ta}{t_{\alpha}}

\begin{document}

\title{Measuring the one-particle
excitations of ultracold fermionic atoms\\
by stimulated Raman spectroscopy}

\author{Tung-Lam Dao}
\affiliation{Centre de Physique Th{\'e}orique, CNRS UMR 7644,
Ecole Polytechnique, Route de Saclay, 91128 Palaiseau Cedex, France}
\author{Antoine Georges}
\affiliation{Centre de Physique Th{\'e}orique, CNRS UMR 7644,
Ecole Polytechnique, Route de Saclay, 91128 Palaiseau Cedex, France}
\author{Jean Dalibard}
\affiliation{Laboratoire Kastler-Brossel, Ecole Normale
Sup\'erieure, 24, rue Lhomond, 75231 Paris Cedex 05, France}
\author{Christophe Salomon}
\affiliation{Laboratoire Kastler-Brossel, Ecole Normale
Sup\'erieure, 24, rue Lhomond, 75231 Paris Cedex 05, France}
\author{Iacopo Carusotto}
\affiliation{CNR-INFM BEC Center and universit\`a di Trento, 38050 Povo, Italy}

\begin{abstract}
We propose a Raman spectroscopy technique which is able to probe
the one-particle Green's function, the Fermi surface, and the quasiparticles
of a gas of strongly interacting ultracold atoms. We give
quantitative examples of experimentally accessible spectra. The
efficiency of the method is validated by means of simulated images
for the case of a usual Fermi liquid as well as for more exotic states:
specific signatures of e.g. a d-wave pseudo-gap are  clearly visible.
\end{abstract}

%\date{\today}
\date{November 6, 2006}
\maketitle

The remarkable advances in handling ultra-cold atomic gases have given
birth to the new field of ``condensed matter physics with light and atoms''.
Key issues in the physics of strongly correlated quantum systems
can be addressed from a new perspective in this context.
The observation of the Mott transition of bosons
in optical lattices~\cite{greiner_mott_nature_2002},
of the superfluidity of fermionic gases~\cite{Superfluid},
%~\cite{greiner_bec_mol_nature_2003,jochim_bec_mol_science_2003,
%zwierlein_bec_mol_prl_2003,bourdel_bec_bcs_prl_2004}
and the recent imaging of Fermi
surfaces~\cite{kohl_fermisurface_prl_2005}
have been important milestones in this respect.
Ultimately fermionic
atoms in optical lattices \cite{bloch_review_natphys_2005,chin}
could help understanding some outstanding problems
of condensed matter physics, such as high-temperature
superconductivity. In this context, a key issue is the nature of
the low-energy excitations of low-dimensional strongly interacting
Fermi systems. There is abundant experimental evidence that those
are highly unconventional, departing from standard Fermi-liquid
theory.

In this letter, we study how to probe the one-particle excitations of
interacting ultracold fermionic atoms using stimulated Raman spectroscopy.
This technique has been considered previously in the context of cold atomic gases,
as an outcoupling technique to produce an atom laser~\cite{hagley_atomlaser_science_1999},
and also as a measurement technique for
bosons~\cite{japha_stimulated_prl_1999,luxat_tunneling_pra_2002,
blakie_raman_2005,Mazets} and fermions~\cite{Zoller,yi_raman_2006}.
Here, we demonstrate that this technique provides,
for strongly-interacting fermion gases, a momentum-resolved access
to key properties of the quasiparticle excitations, such as their
dispersion relation and lifetime.
%\cs {\itSentence below rewritten [JEAN]:}
It also allows for a determination of the Fermi surface itself
in strongly interacting regimes, whereas previously demonstrated
methods~\cite{kohl_fermisurface_prl_2005} apply to the non-interacting
case. Furthermore, it is shown that the suppression
of quasiparticles due to a pseudogap in the excitation spectrum
can also be detected by this method.

In a conventional Fermi liquid, low-energy excitations are built
out of quasiparticles~\cite{mbt_books}. Those are
characterized by their dispersion relation, i.e the energy $\xik$
(measured from the ground-state energy) necessary to create such
an excitation with (quasi-) momentum $\bk$. The interacting system
possesses a Fermi surface (FS) defined by the location in
momentum space on which the excitation energy vanishes:
$\xi_{\kF}=0$. Close to a given point on the FS, the
quasiparticle energy vanishes as:
$\xik\sim\vF(\kF)\cdot(\bk-\kF)+\cdots$, with $\vF$ the local
Fermi velocity (inversely related to the effective mass).
Quasiparticle excitations have a finite lifetime $\Gk^{-1}$, and
are well defined provided $\Gk$ vanishes faster than $\xik$ as the
FS is approached ($\Gk\sim\xik^2$ in Fermi liquid theory). In
contrast, one-particle excitations in the ``normal'' (i.e
non-superconducting) state of the cuprate superconductors (SC)
reveal strong deviations from this
behaviour~\cite{damascelli_rmp_2003}. Reasonably well-defined
quasiparticle excitations only exist close to the diagonal
direction of the Brillouin zone (the ``nodal'' direction
along which the $d$-wave gap vanishes in the SC phase), and even
there $\Gk$ is rather large. Away from this direction (in the
``antinodal'' region), excitations appear to be short-lived and
gapped already above the SC critical temperature (the so-called
pseudogap phenomenon). This momentum-space differentiation is a
key to the physics of cuprates.
% \cs {\it Theory references removed from here}
%In fact, recent theoretical
%investigations~\cite{cuprates} by a variety of methods strongly
%suggest that this phenomenon is present in simple theoretical
%models such as the two-dimensional Hubbard model.

Experiments probing directly non-diagonal one-particle correlators
$\langle\psi^\dagger(\br, t)\psi(\br', t')\rangle$ of a
many-body system are therefore highly desirable but also
relatively scarce. Most physical measurements indeed provide
information on two-particle correlators of the form
$\langle\psi^\dagger(\br,t)\psi(\br,t)\,
\psi^\dagger(\br',t')\psi(\br',t')\rangle$~\footnote{Information
on one-particle correlators of a Bose system
can be obtained from two-particle ones,
when the system is made to interfere with either another identical system
%from interference experiments with another identical system~
\cite{PAD} or with a reference condensate~\cite{NCK}.}.
%
%\bibitem{PAD} A. Polkovnikov, E. Altman, E. Demler, PNAS {\bf 103}, 6125
%(2006)
%\bibitem{NCK} Q. Niu, I. Carusotto, A. B. Kuklov, Phys. Rev. A 73,
%053604 (2006)
%
Examples are neutron scattering or transport measurements in the
solid-state context~\cite{mbt_books}, and Bragg
scattering~\cite{carusotto_bragg_jpb_2006} or noise correlations
measurements~\cite{altman_noise_pra_2004} in the cold atom
context. For Bose systems with a finite condensate density $n_0$,
the two-particle correlator is closely related to the one-particle
correlator via terms such as $n_0\,\langle\psi^\dagger(\br,
t)\,\psi(\br',t')\rangle$.
%v2
By contrast, in Fermi systems, the distinction between
one- and two-particle correlators is essential
and specific measurement techniques of the former are requested.

In solids, angle-resolved photoemission spectroscopy (ARPES)
provides a direct probe of the one-particle
spectrum~\cite{damascelli_ARPESintro_physscripta_2004}, and has played a key role in
revealing momentum-space differentiation in
cuprates~\cite{damascelli_rmp_2003}. It consists in measuring the
energy and momentum of electrons emitted out of the solid exposed
to an incident photon beam. In the simplest approximation, the
emitted intensity can be related to the single-electron spectral
function, defined at $T=0$ and for $\omega<0$, i.e for
hole-like excitations by:
$A(\bk,\omega)=\sum_n |\langle
\Psi_n^{N-1}|\psi_{\bk}|\Psi_0^N\rangle|^2\,
\delta(\omega+\mu+E_n-E_0)$. In this expression, $\psi_\bk$ is
a destruction operator for an electron with momentum $\bk$,
$\Psi_0^N$ is the ground-state of the N-particle system and
$\Psi_n^{N-1}$ are the eigenstates of the system with one less
particle. In a conventional Fermi liquid, and for momenta close to
the FS, the spectral function can be separated~\cite{mbt_books}
into a coherent quasiparticle contribution and an incoherent contribution
%(Fig.~\ref{fig:raman})
: $A=A_{\rm{QP}}+A_{\rm{inc}}$, with
$\pi A_{\rm{QP}}(\bk,\omega)\simeq\zk\Gk/[(\omega-\xik)^2+\Gk^2]$
and $A_{\rm{inc}}$ widely
spread in frequency. Only a finite fraction $\zk<1$ of the total
spectral weight corresponds to long-lived quasiparticle
excitations.
%
%\begin{figure}
%\includegraphics[width=6cm]{Figs/Raman.eps}
%%\includegraphics[width=6cm]{Figs/Raman}
%\caption{\label{fig:raman} Raman process: transfer from an
%internal state $\alpha$ to another internal state $\beta$ through
%an excited state $\gamma$. The momentum-resolved spectral function
%is schematized, consisting of a quasiparticle peak and an
%incoherent background. }
%\end{figure}

%v2
In this paper, we consider stimulated Raman spectroscopy
%as a probe of one-particle excitations in
on a two-component mixture of ultracold fermionic atoms in
two internal states $\a$ and $\a'$.
%
%Stimulated Raman spectroscopy has been considered previously in
%the context of cold atomic gases, both as an outcoupling technique
%to produce an atom laser~\cite{hagley_atomlaser_science_1999} and
%as a measurement technique for
%bosons~\cite{japha_stimulated_prl_1999,luxat_tunneling_pra_2002,
%blakie_raman_2005,Mazets} and
%fermions~\cite{Zoller,yi_raman_2006}.
%In the Raman process
%of Fig.~\ref{fig:raman}
%In the stimulated Raman process,
Atoms are transferred from $\a$ into
another internal state $\b\neq\a,\a'$, through an intermediate
excited state $\g$, using two laser beams of wavectors
$\bk_{1,2}$ and frequencies $\omega_{1,2}$. If $\omega_1$ is
sufficiently far from single photon resonance to the excited $\g$
state, we can neglect spontaneous emission. Eliminating the
excited state, we write an effective hamiltonian: $\hat{V}=
C\int d\br\, \psida(\br)\psib(\br) e^{i(\bk_1-\bk_2)\cdot\br}
a^\dagger_1 a_2+\mathrm{h.c.}$, in which $a^\dagger_1$ ($a_2$)
denotes the creation (destruction) operator of a photon
respectively in mode $1$ ($2$) and the constant $C$ is
proportional to the product of the dipole matrix elements
$d_{\a\g}$ and $d_{\b\g}$ of the optical transitions and inversely
proportional to the detuning from the excited state.

The total transfer rate to state $\b$ can be
calculated~\cite{japha_stimulated_prl_1999,luxat_tunneling_pra_2002,blakie_raman_2005}
using the Fermi golden rule:
\begin{multline}
R(\Dk,\Dom)=
|C|^{2}n_{1}(n_{2}+1)\int_{-\infty}^{\infty}\!\!\!\!\!dt \int
\!d\br\,d\br'\,e^{i[\,\Dom\,t
-\Dk\cdot(\br-\br')]} \\
\times\, g_\b(\br,\br';t)
\langle \psi^{+}_{\alpha}(\br,t)\psi_{\alpha}(\br',0) \rangle.
\label{eq:rate1}
\end{multline}
Here $\Dk=\bk_1-\bk_2$ and $\Dom=\omega_1-\omega_2+\mu$ with $\mu$
the chemical potential of the interacting gas, and $n_{1,2}$
the photon numbers present in the laser beams. Assuming that no
atoms are initially present in $\b$ and that the scattered atoms
in $\b$ do not interact with the atoms in the initial $\a,\a'$
states, the free propagator for $\b$-state atoms in vacuum is to
be taken: $g_\b(\br,\br';t)\equiv \langle
0_{\b}|\,\psib(\br,t)\psidb(\br',0)|0_{\b}\rangle$. The
correlation function entering (\ref{eq:rate1}) is proportional to
the one-particle Green's function ~\footnote{The superscript $<$
indicates that $\psi^\dagger$ is always to the left of $\psi$.
%independently of the sign of $t$.
Operators are evolved in the grand-canonical ensemble~\cite{mbt_books}.}
$\langle
\psi^{+}_{\alpha}(\br,t)\psi_{\alpha}(\br',0) \rangle
= -i G_{\alpha}^{<}\,(\br',\br,-t)$ of the strongly-
interacting Fermi system.
%and thus carries information on its many-body density matrix.
%
%A complete experimental determination of the one-particle Green's function
%$G_{\alpha}^{<}$  would require to measure the rate $R(\Dk,\Dom)$
%for many different values of $\Dk$ and $\Dom$ and to perform
%an inverse Fourier transform. This being a cumbersome task, it
%is useful to identify ways of extracting useful information on the
%many-body system from a limited number of measurements.
%
For a uniform system,
%uniform lattice case with a hopping matrix element $t$.
the rate (\ref{eq:rate1}) can be
related to the spectral function $A(\bk,\omega)$ of atoms in the
internal state $\a$ by~\cite{luxat_tunneling_pra_2002}:
\begin{equation}
R(\Dk,\Dom)\propto\,
%\frac{|C|^{2}n_{1}n_{2}}{(2\pi)^{3}}
\int \! d\bk\, n_{F}\,(\varepsilon_{\bk\beta}-\Dom)\,
A(\bk-\Dk,\varepsilon_{\bk\beta}-\Dom)
\label{eq:rate2}
\end{equation}
in which the Green's function has been expressed in terms of
the spectral function and the Fermi factor $n_F$ as~\cite{mbt_books}:
$G_{\alpha}^{<}\,(\bk,\omega)=i\,n_F(\omega)\,A(\bk,\omega)$.
%As usual in the solid state literature, single-particle energies
%$\xik^0=\varepsilon_{\bk\alpha}-\mu$ for the $\a$ state (as well
%as the frequency $\omega$) are here measured from the chemical
%potential, i.e the Fermi energy, at $T=0$.
%The same choice is made for the frequency $\omega$.

%%%%%%%%%%%%%%%%%%%%%%%%%%%%%%%%%%%%%%
%Analysis of rate formula and various models for spectral function
%%%%%%%%%%%%%%%%%%%%%%%%%%%%%%%%%%%%%%
%
%Homogeneous case, free system and Fermi liquid.
%
%\cs {\it [Paragraph split in two and reorganized by popular demand !
%I have changed the presentation a bit, in fact: it was not very clever to
%present this as an analysis of the non-interacting system first: rather we can
%present it as neglecting first the QP lifetime, and incoherent part and then
%looking at increasingly complex cases}
%
In order to physically understand which information can be
extracted from a measurement of the rate (\ref{eq:rate2}),
let us first approximate the spectral function by
$A(\bk,\omega)=\delta(\omega-\xik)$, i.e neglect the
incoherent part and consider quasiparticles with an infinite lifetime.
%$\xik^0=\varepsilon_{\bk\alpha}-\mu$ the single-particle energies
%for the $\a$ state (counted from the Fermi energy).
The Raman rate then reads at $T=0$:
$R=\int_{\xik<0} d\bk\,\delta(\varepsilon_{\bk+\Dk,\beta}-\xik-\Dom)$.
Contributions to
this integral come from momenta inside the FS ($\xik<0$) which
satisfy the Raman resonance condition
$\varepsilon_{\bk+\Dk,\beta}-\xik=\Dom$. When the frequency
shift $\Omega$ is small, $R$ vanishes since there is no
available phase-space satisfying these constraints.
%
%The (non-interacting) scattered $\beta$ atoms have a parabolic
%dispersion $\ekb=\eb+\bk^2/2M$, with a mass $M$ equal to either
%the bare atomic mass, or the effective mass at the lower band edge
%if the $\beta$ atoms also feel the optical lattice.
%
The smallest frequency at which a signal starts to be observed is
$\Dom_T=\rm{Min}_{\bk}\ekb\equiv\eb$~\footnote{The threshold
corresponds to $(\o_1-\o_2)_T=\eb-\mu$. This is of order $\eb-\ea$
at weak coupling, while interactions can lead to a shift
comparable to the Hubbard coupling.}. This corresponds to a
momentum transfer $\Dk=-\bk_F$ which lies itself on the FS (i.e
$\xi_{\mathbf{k}_F}=0$)~\footnote{The accessible range of $\Dk$ is
$0\leq c|\Dk|/ 2 \leq \omega_1\simeq \omega_2$. In a red-detuned
lattice with respect to $\omega_{1,2}$, all values of $\Dk$ in the
first Brillouin zone are therefore accessible.}.
For $\Omega$ very close to the
extinction threshold ($\Delta\Dom=\Dom-\DomT\gtrsim 0$),
the region in momentum space inside which a sizeable
transfer rate $R$ is measured consists of
a shell surrounding the FS, centered at
a momentum $\bq$ such that $\Delta\Omega=-\xi_{-\bq}\sim
\vF(\kF)\cdot(\bq+\kF)$, and of width
$\Delta q_{\parallel}\sim\sqrt{2M\,\Delta\Dom}$.
In these expressions, $M$ is the effective
mass at the bottom of the $\beta$-band and $\vF$ is the Fermi
velocity.

This analysis remains unchanged when considering quasiparticles
with a finite lifetime $\Gamma^{-1}$ (uniform along the FS),
the width of the momentum shell being simply replaced by
$\Delta q_{\parallel}\sim\sqrt{2M\,\Delta\Dom}+\Gamma/v_F$.
Hence, measuring the Raman signal for $\Dom$ close to the extinction
threshold $\Dom_T$ and sweeping over $\Dk$, provides a determination of the FS
in an interacting system
(while the method of~\cite{kohl_fermisurface_prl_2005} applies
to non-interacting fermions). It also gives access to the
velocity of quasiparticles (from the displacement of the measured signal
as a function of $\Omega$) and to their lifetime (from the width
of the momentum-shell).

\begin{figure}[tb]
\includegraphics[width=7cm]{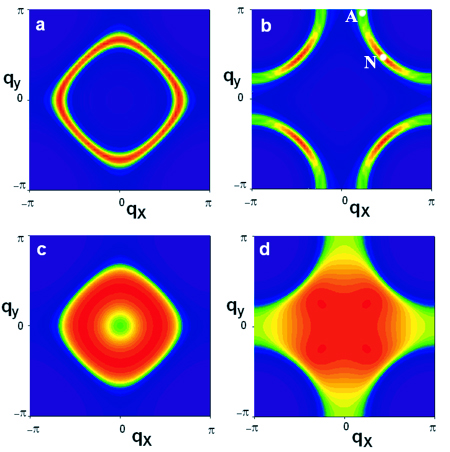}
\caption{Intensity plots of the Raman rate $R(\Dk,\Dom)$, for
$\Dom$ close to threshold $\DomT$ ($\Delta\Dom=0.01\ta$). (a)
Non-interacting fermions on the homogeneous 2D square lattice with
density
 $n_\a=0.22$ and a Lorentzian broadening of the spectral function
 $\Gamma=0.4 \ta$ uniform in $\bk$-space.
(b) Model $d$-wave pseudogap
state (see text), with $\Delta_0=0.1\,\ta$, $\Gamma_0=0.05\,\ta,
\Gamma_1=0.4\,\ta$. The plot is for a hole-doped system
($n_\a=0.45$) with a nearest ($\ta$) and next-nearest neighbour
($\ta^\prime$) hopping,
%on the square lattice,
with $\ta^\prime/\ta=-0.3$ (typical for cuprates, but similar effects are
expected also for smaller $|\ta^\prime/\ta|$).
(c) and (d): same as (a) and (b) in the
presence of a harmonic trap ($\omt=0.02\ta$).
The pseudogap and nodal-antinodal differentiation are clearly visible
in both 1b and 1d.
\label{fig:FS_q-maps} }
\end{figure}

\begin{figure}[tb]
  \includegraphics[width=9cm]{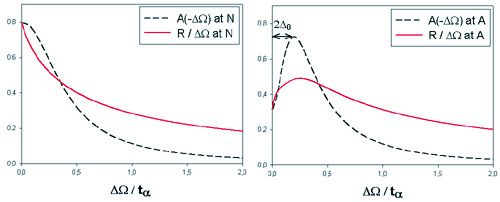}\\
  \caption{Comparison between the
  spectral function $A$ and the Raman rate $R/\Delta\Omega$
  for two points in momentum-space indicated on Fig.~1.
  In the nodal direction (N; left), the spectrum displays a
  quasiparticle peak, while in the antinodal direction
  (A; right) a depletion of the signal is observed at low energy,
  corresponding to the pseudogap.}
\label{Fig:Comparison}
\end{figure}

Examples of numerically simulated spectra for
uniform interacting systems are given in
Fig.~\ref{fig:FS_q-maps}a-b, where a color intensity plot of the Raman
rate (\ref{eq:rate2}) is shown for a fixed value of the frequency shift
close to threshold.
%%
% Model forms for hi-Tc
%%
In (a), we consider the case of a
%uniformly broadened
Lorentzian spectral function centered around the free dispersion relation of
a two-dimensional square lattice: $\xik=-2\ta(\cos k_x + \cos
k_y)-\mu$. In (b) a phenomenological
form~\cite{norman_pheno_prb_1998} of the spectral function is
used,
%v2 shorten
which captures the main aspects of the ARPES
data in the non-SC (``normal'') state of high-temperature
superconductors.
%v2 Insist on pseudo-gap, not SC gap w/ LRO
The key feature entering
this phenomenological form is a {\it pseudo-gap} with $d$-wave symmetry
$\Delta_\bk=\Delta_0 (\cos k_x - \cos k_y)$,
corresponding to a depletion of low-energy excitations even when
no long-range SC order is present.
$\Delta_\bk$ vanishes along
the zone diagonal (nodes) and is maximum
along $(0,0)-(\pi,0)$ (antinodes).
A self-energy $\Delta_\bk^2/(\omega+\xik)$ is a
convenient modelization of this effect. In addition, finite
lifetimes effects are introduced, resulting in the form:
$\pi A(\bk,\omega)=-\rm{Im}\left[\omega-\xik+i\Gamma_1-
\Delta_\bk^2\,/(\omega+\xik+i\Gamma_0)\right]^{-1}$. This
corresponds to a quasiparticle dispersion which is gapped out
except at the nodes. The width
$\Gamma_\bk=\Gamma_1+\Delta_\bk^2\,\Gamma_0/[(\omega+\xik)^2+\Gamma_0^2]$
is largest near the antinodes. This
form of $A(\bk,\omega)$  also provides a reasonable qualitative
description of recent theoretical results for the two-dimensional
Hubbard model~\cite{cuprates}.
The momentum space differentiation
encoded in the model spectral function is clearly visible
%v2 shorten
%from the Raman intensity plot of
on Fig.~\ref{fig:FS_q-maps}b, with nodal regions displaying
quasiparticles while antinodal ones are gapped
out and short-lived. This illustrates how the Raman spectroscopy
method can be used to determine the FS not only of a Fermi liquid
but also of a strongly interacting system
%v2
with suppressed quasiparticles.
%
% This in contrast
%to the experimental techniques pioneered in
%\cite{kohl_fermisurface_prl_2005}, which
%%are based on a ballistic expansion and are therefore limited
%apply to non-interacting systems.
%
%v2 New figure and comparison between R and A
In Fig.~\ref{Fig:Comparison} we further show that the lineshape of
the Raman signal for a fixed value of $\Dk$ does reveal the essential
features of the spectral function,
namely quasiparticles at the nodes and a pseudo-gap at the antinodes.

%%
% Effect of the trapping potential
%%
%v2 shorten
%Most cold atom experiments are performed within a (generally
%harmonic) trap. Thus, it is important to include the effect of the
%spatial inhomogeneities and verify that it does not spoil the
%predicted signatures.
%
Since most cold atom experiments are performed in a trap,
it is important to verify that the spatial inhomogeneity does
%we now verify that the inhomogeneity does
not spoil the predicted signatures.
%
%The simplest way of taking into account the
%trapping potential is to use a local density approximation, where
%the observed signal is the sum of the contributions of the
%different points $\bR$ of the trap. Spatial inhomogeneity enters
%via the fact that the local chemical potential has a different
%value at different spatial positions, $\mu_\bR=\mu-M\omt^2\bR^2/2$.
%
Within the local density approximation, the observed signal is
the sum of the contributions of the different points $\bR$ of the trap,
with a local chemical potential $\mu_\bR=\mu-M\omt^2\bR^2/2$.
The results are summarized in
Figs.~\ref{fig:FS_q-maps}(c,d) for physical situations such that
the value of the chemical potential at the trap center coincides with
that of the homogeneous system in Figs.\ref{fig:FS_q-maps}(a,b).
As expected, the intensity map is
now a superposition of the Fermi surfaces corresponding to all the
densities realized in the trap. The outer shell delimited by the
extinction of the signal still gives a direct access to the FS
corresponding to the highest densities at the centre of the trap.
The typical signatures of an unconventional state remain clearly visible in
the trap as well: in Fig.~\ref{fig:FS_q-maps}d, the
nodal-antinodal differentiation is apparent in the outer shell of
%\cs {\it [Clarification following CS's comment]}
this plot, as seen from the suppressed intensity along
the antinodal direction.
A possible way
of revealing the region around the Fermi surface is to measure the
intensity maps for two, slightly different values of the frequency
and/or the total atom number, and then take their difference: the
resulting differential images for the trapped system (not shown)
recover the same qualitative features of the homogeneous system
shown in Fig.~\ref{fig:FS_q-maps}a-b.

\begin{figure}[tb]
\includegraphics[width=7.5cm]{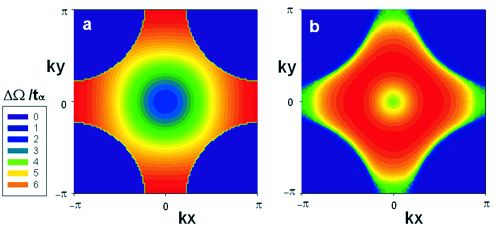}
\caption{(a) Colour plot illustrating the selective addressing of
$\bk$-space by a proper choice of $\Omega$ (cf. colour scale on the
left). (b) Time of flight
$\bk$-map obtained by integrating the Raman intensity for
$\Delta\Dom=\Dom-\Omega_{T}$ varied in the range
$[2.4\ta,6.8\ta]$. The dispersion relation of the $\beta$-atoms is
taken as $\ekb=\eb-2t_{\b}(2+\cos k_x+\cos k_y)$ with
$t_{\b}=1.5\ta$ (note that interactions will renormalize downwards
the effective $\ta$ even if bare values are equal). Parameters are
as in Fig.1c and $\Dk=0$. \label{fig:TOF}}
\end{figure}

%%%%%%%%%%%%%%%%%%%%%%%%%%%%%%%%%%%%%%%%%%%%%%%%%%%%%%%%%%%%%%%
% TOF measurements and k-maps of intensity
%%%%%%%%%%%%%%%%%%%%%%%%%%%%%%%%%%%%%%%%%%%%%%%%%%%%%%%%%%%%%%%
The discussion so far has assumed that it is possible to repeat
the measurement of the total rate $R$ for several different values of
$\Dk$. In some cases, a different scheme with a momentum-selective
detection of  the scattered $\beta$ atoms may be instead
favorable, quite similar to ARPES in solids.
A single value of $\Dk$ is used, and a time of
flight expansion of the $\b$ atoms cloud is performed
(after suddenly turning
off the trap and the lattice potential) in order to reconstruct the
momentum distribution of the atoms.
As shown in Fig.~\ref{fig:TOF}a, the
Raman resonance condition allows for a selective addressing of the
different regions in $\bk$ by tuning the frequency $\Omega$.
The number of Raman-scattered atoms with final momentum $\bk$ is proportional to
the integrand $n_{F}(\varepsilon_{\bk\beta}-\Dom)\,A(\bk-\Dk,\varepsilon_{\bk\beta}-\Dom)$
of (\ref{eq:rate2}).
Fig.3b shows that the resulting $\bk$-space intensity map is able to
reveal the details of the pseudogap physics, in particular its
$\bk$-dependence.
%\cs
By varying both $\Dom$ and $\Dk$, Raman scattering offers more
possibilities for probing the system in a momentum-selective way
than microwave spectroscopy techniques~\cite{Grimm}.

%%%%%%%%%%%%%%%%%%%%%%%%%%%%%%%%%%%%%%%%%%%%%%%%%%%%%%%%%%%%%%%
% Discussion of feasibility of experiment: which hyperfine levels to choose,
% orders of magnitude, effect of confining potential, etc...
%%%%%%%%%%%%%%%%%%%%%%%%%%%%%%%%%%%%%%%%%%%%%%%%%%%%%%%%%%%%%%%%
%
%\cs{\it Changed w/ hel[p of CS}
%
As a final point, we discuss some orders of magnitude which are
important for the actual feasibility of the experiments proposed
in this article. Specifically, we consider $^6$Li atoms
(see also Ref.~\cite{yi_raman_2006})
%where high-temperature superfluidity has been observed using
in the two lowest hyperfine states $|\alpha\rangle= |I_z=1,m=-1/2\rangle$ and
$|\alpha'\rangle=|0,-1/2\rangle$.
The coupling between these two states can be made
very large thanks to the Feshbach resonance at $834$~Gauss. On the
other hand, if we choose the final state of the Raman process to be
$|\beta\rangle=|1,1/2\rangle$ with the same nuclear spin component as $\alpha$,
the interaction of $|\beta\rangle$ with both $|\alpha\rangle$ and $|\alpha'\rangle$ is
non-resonant, corresponding to a low value of the scattering length
$a_{\alpha,\beta}\simeq a_{\alpha',\beta}\simeq 2.5\,$
nm~\cite{vanabeelen_li6_pra_1997.pdf}.
This yields a typical scale for the interaction energy between an
atom in $\beta$ and the background in $\alpha,\alpha'$ which is smaller
than the typical bandwidth, and hence negligible.
Furthermore, taking typical values
for the lattice wavelength $\lambda\sim
800\rm{nm}$, the atomic density $\rho\sim (2/\lambda)^3$ and the
recoil velocity $v\sim h/(M\lambda)$ of atoms in state $\beta$, we
evaluate the collision rate to be $\gamma_{\rm c}= \rho \sigma
v\sim 10^2$\,s$^{-1}$. The Raman detection sequence can therefore
be performed in a time scale of the order of a few milliseconds,
yielding an energy resolution in the $100\rm{Hz}$ range. Losses
due to inelastic transitions from state $\beta$ have a rate $\sim
10^{-12} \rm{cm}^3\rm{s}^{-1}$ and can be neglected on this time
scale.

In summary, we have proposed a Raman spectroscopy technique which,
analogously to ARPES in solid-state physics, is able to probe
the one-body Green's function. This technique can be used to obtain
information on the Fermi surface, and on the quasiparticles
%v2
(or absence thereof)
of a gas of fermionic atoms, even in strongly-correlated states. In the
near future, this technique may play an important role in the
experimental characterization of the novel quantum states of
matter that can be obtained with ultracold atoms in optical
lattices.

\acknowledgments We are grateful to T.~Esslinger for an
interesting discussion. We acknowledge support from the ANR under
contract ``GASCOR'', from IFRAF, and from CNRS and Ecole
Polytechnique. Laboratoire Kastler Brossel is a research unit of
Ecole normale sup\'erieure and Universit\'e Paris 6, associated to
CNRS.

%\end{thebibliography}


\begin{thebibliography}{24}
\expandafter\ifx\csname
natexlab\endcsname\relax\def\natexlab#1{#1}\fi
\expandafter\ifx\csname bibnamefont\endcsname\relax
  \def\bibnamefont#1{#1}\fi
\expandafter\ifx\csname bibfnamefont\endcsname\relax
  \def\bibfnamefont#1{#1}\fi
\expandafter\ifx\csname citenamefont\endcsname\relax
  \def\citenamefont#1{#1}\fi
\expandafter\ifx\csname url\endcsname\relax
  \def\url#1{\texttt{#1}}\fi
\expandafter\ifx\csname
urlprefix\endcsname\relax\def\urlprefix{URL }\fi
\providecommand{\bibinfo}[2]{#2}
\providecommand{\eprint}[2][]{\url{#2}}

\bibitem[{\citenamefont{Greiner et~al.}(2002)\citenamefont{Greiner, Mandel,
  Esslinger, H\"ansch, and Bloch}}]{greiner_mott_nature_2002}
\bibinfo{author}{\bibfnamefont{M.}~\bibnamefont{Greiner et~al.}},
  \bibinfo{journal}{Nature} \textbf{\bibinfo{volume}{415}}, \bibinfo{pages}{39}
  (\bibinfo{year}{2002}).

\bibitem[{\citenamefont{Greiner et~al.}(2003)\citenamefont{Greiner, Regal, and
  Jin}}]{Superfluid}
%\cs {\it Reference to Kinast et al. added [CS]}
  See, for example:
\bibinfo{author}{\bibfnamefont{M.}~\bibnamefont{Greiner et~al.}},
  \bibinfo{journal}{Nature} \textbf{\bibinfo{volume}{537}},
  \bibinfo{pages}{426} (\bibinfo{year}{2003});
\bibinfo{author}{\bibfnamefont{S.}~\bibnamefont{Jochim et~al.}},
  \bibinfo{journal}{Science} \textbf{\bibinfo{volume}{302}},
  \bibinfo{pages}{2101} (\bibinfo{year}{2003});
\bibinfo{author}{\bibfnamefont{M.~W.} \bibnamefont{{Zwierlein} et~al.}},
  \bibinfo{journal}{Phys. Rev. Lett.} \textbf{\bibinfo{volume}{91}},
  \bibinfo{pages}{250401} (\bibinfo{year}{2003});
\bibinfo{author}{\bibfnamefont{J.}~\bibnamefont{{Kinast} et~al.}},
  \bibinfo{journal}{Phys. Rev. Lett.} \textbf{\bibinfo{volume}{92}},
  \bibinfo{pages}{150402} (\bibinfo{year}{2004});
\bibinfo{author}{\bibfnamefont{T.}~\bibnamefont{{Bourdel} et~al.}},
  \bibinfo{journal}{Phys. Rev. Lett.} \textbf{\bibinfo{volume}{93}},
  \bibinfo{pages}{050401} (\bibinfo{year}{2004});

\bibitem[{\citenamefont{{K{\" o}hl} et~al.}(2005)\citenamefont{{K{\" o}hl},
  {Moritz}, {St{\" o}ferle}, {G{\" u}nter}, and
  {Esslinger}}}]{kohl_fermisurface_prl_2005}
\bibinfo{author}{\bibfnamefont{M.}~\bibnamefont{{K{\" o}hl} et~al.}},
  \bibinfo{journal}{Phys. Rev. Lett.} \textbf{\bibinfo{volume}{94}},
  \bibinfo{pages}{080403} (\bibinfo{year}{2005}).

\bibitem[{\citenamefont{Bloch}(2005)}]{bloch_review_natphys_2005}
For a review, see:
\bibinfo{author}{\bibfnamefont{I.}~\bibnamefont{Bloch}},
  \bibinfo{journal}{Nature Physics} \textbf{\bibinfo{volume}{1}},
  \bibinfo{pages}{24} (\bibinfo{year}{2005}).

\bibitem{chin} J. K. Chin et al., Nature {\bf 443}, 961 (2006).

%v2
\bibitem[{\citenamefont{Hagley et~al.}(1999)\citenamefont{Hagley, Deng, Kozuma,
  Wen, Helmerson, Rolston, and Phillips}}]{hagley_atomlaser_science_1999}
\bibinfo{author}{\bibfnamefont{E.~W.} \bibnamefont{Hagley et~al.}},
  \bibinfo{journal}{Science}
  \textbf{\bibinfo{volume}{283}}, \bibinfo{pages}{1706} (\bibinfo{year}{1999}).

\bibitem[{\citenamefont{{Japha} et~al.}(1999)\citenamefont{{Japha}, {Choi},
  {Burnett}, and {Band}}}]{japha_stimulated_prl_1999}
\bibinfo{author}{\bibfnamefont{Y.}~\bibnamefont{{Japha} et~al.}},
  \bibinfo{journal}{Phys. Rev. Lett.}
  \textbf{\bibinfo{volume}{82}}, \bibinfo{pages}{1079} (\bibinfo{year}{1999}).

\bibitem[{\citenamefont{{Luxat} and
  {Griffin}}(2002)}]{luxat_tunneling_pra_2002}
\bibinfo{author}{\bibfnamefont{D.~L.} \bibnamefont{{Luxat}}} \bibnamefont{and}
  \bibinfo{author}{\bibfnamefont{A.}~\bibnamefont{{Griffin}}},
  \bibinfo{journal}{Phys. Rev. A} \textbf{\bibinfo{volume}{65}},
  \bibinfo{pages}{043618} (\bibinfo{year}{2002}).

\bibitem[{\citenamefont{{Blair Blakie}}(2005)}]{blakie_raman_2005}
\bibinfo{author}{\bibfnamefont{P.}~\bibnamefont{{Blair Blakie}}}, \eprint{cond-mat/0508365}.

\bibitem{Mazets}
I. E. Mazets, G. Kurizki, N. Katz, and N. Davidson, Phys. Rev.
Lett. {\bf 94} 190403 (2005).

\bibitem{Zoller}
P. T\"orm\"a and P. Zoller, Phys. Rev. Lett. \textbf{85}, 487
(2000).

\bibitem[{\citenamefont{{Yi} and {Duan}}(2006)}]{yi_raman_2006}
\bibinfo{author}{\bibfnamefont{W.}~\bibnamefont{{Yi}}} \bibnamefont{and}
  \bibinfo{author}{\bibfnamefont{L.~} \bibnamefont{{Duan}}},
  \eprint{cond-mat/0605440}.


\bibitem[{\citenamefont{Abrikosov et~al.}(1963)\citenamefont{Abrikosov, Gorkov,
  and Dzyaloshinski}}]{mbt_books}
\bibinfo{author}{\bibfnamefont{A.~A.} \bibnamefont{Abrikosov et~al.}},
  \emph{\bibinfo{title}{Methods of Quantum Field
  Theory in Statistical Physics}} (\bibinfo{publisher}{Dover},
  \bibinfo{address}{New York}, \bibinfo{year}{1963});
%\bibitem[{\citenamefont{Mahan}(1981)}]{mahan_book}
\bibinfo{author}{\bibfnamefont{G.~D.} \bibnamefont{Mahan}},
  \emph{\bibinfo{title}{Many Particle Physics}} (\bibinfo{publisher}{Plenum},
  \bibinfo{address}{New York}, \bibinfo{year}{1981}).


\bibitem[{\citenamefont{{Damascelli} et~al.}(2003)\citenamefont{{Damascelli},
  {Hussain}, and {Shen}}}]{damascelli_rmp_2003}
\bibinfo{author}{\bibfnamefont{A.}~\bibnamefont{{Damascelli} et~al.}},
  \bibinfo{journal}{Rev. Mod. Phys.}
  \textbf{\bibinfo{volume}{75}}, \bibinfo{pages}{473}
  (\bibinfo{year}{2003}).

%\bibitem[{\citenamefont{Mahan}(1981)}]{mahan_book}
%\bibinfo{author}{\bibfnamefont{G.~D.} \bibnamefont{Mahan}},
%  \emph{\bibinfo{title}{Many Particle Physics}} (\bibinfo{publisher}{Plenum},
%  \bibinfo{address}{New York}, \bibinfo{year}{1981}).

\bibitem[{\citenamefont{Carusotto}(2006)}]{carusotto_bragg_jpb_2006}
D. M. Stamper-Kurn, {\em et al.}, Phys. Rev. Lett. {\bf 83}, 2876 (1999);
\bibinfo{author}{\bibfnamefont{I.}~\bibnamefont{Carusotto}},
  \bibinfo{journal}{J. Phys. B: At.Mol.Opt.Phys.}
  \textbf{\bibinfo{volume}{39}}, \bibinfo{pages}{S211} (\bibinfo{year}{2006}).


\bibitem[{\citenamefont{{Altman} et~al.}(2004)\citenamefont{{Altman}, {Demler},
  and {Lukin}}}]{altman_noise_pra_2004}
S. F\"olling {\it et al.}, Nature {\bf 434}, 481-484
(2005); M. Greiner, C. A. Regal, J. T. Stewart, and D. S. Jin,
Phys. Rev. Lett. {\bf 94}, 110401 (2005);
\bibinfo{author}{\bibfnamefont{E.}~\bibnamefont{{Altman} et~al.}},
  \bibinfo{journal}{Phys. Rev. A} \textbf{\bibinfo{volume}{70}},
  \bibinfo{pages}{013603} (\bibinfo{year}{2004}).

\bibitem[{\citenamefont{{Damascelli}}(2004)}]{damascelli_ARPESintro_physscript%
a_2004}
\bibinfo{author}{\bibfnamefont{A.}~\bibnamefont{{Damascelli}}},
  \bibinfo{journal}{Physica Scripta Volume T} \textbf{\bibinfo{volume}{109}},
  \bibinfo{pages}{61} (\bibinfo{year}{2004}).

%\bibitem[{\citenamefont{Hagley et~al.}(1999)\citenamefont{Hagley, Deng, Kozuma,
%  Wen, Helmerson, Rolston, and Phillips}}]{hagley_atomlaser_science_1999}
%\bibinfo{author}{\bibfnamefont{E.~W.} \bibnamefont{Hagley et~al.}},
%  \bibinfo{journal}{Science}
%  \textbf{\bibinfo{volume}{283}}, \bibinfo{pages}{1706} (\bibinfo{year}{1999}).
%
%\bibitem[{\citenamefont{{Japha} et~al.}(1999)\citenamefont{{Japha}, {Choi},
%  {Burnett}, and {Band}}}]{japha_stimulated_prl_1999}
%\bibinfo{author}{\bibfnamefont{Y.}~\bibnamefont{{Japha} et~al.}},
%  \bibinfo{journal}{Phys. Rev. Lett.}
%  \textbf{\bibinfo{volume}{82}}, \bibinfo{pages}{1079} (\bibinfo{year}{1999}).
%
%\bibitem[{\citenamefont{{Luxat} and
%  {Griffin}}(2002)}]{luxat_tunneling_pra_2002}
%\bibinfo{author}{\bibfnamefont{D.~L.} \bibnamefont{{Luxat}}} \bibnamefont{and}
%  \bibinfo{author}{\bibfnamefont{A.}~\bibnamefont{{Griffin}}},
%  \bibinfo{journal}{Phys. Rev. A} \textbf{\bibinfo{volume}{65}},
%  \bibinfo{pages}{043618} (\bibinfo{year}{2002}).
%
%\bibitem[{\citenamefont{{Blair Blakie}}(2005)}]{blakie_raman_2005}
%\bibinfo{author}{\bibfnamefont{P.}~\bibnamefont{{Blair Blakie}}}, \eprint{cond-mat/0508365}.
%
%\bibitem{Mazets}
%I. E. Mazets, G. Kurizki, N. Katz, and N. Davidson, Phys. Rev.
%Lett. {\bf 94} 190403 (2005).
%
%\bibitem{Zoller}
%P. T\"orm\"a and P. Zoller, Phys. Rev. Lett. \textbf{85}, 487
%(2000).
%
%\bibitem[{\citenamefont{{Yi} and {Duan}}(2006)}]{yi_raman_2006}
%\bibinfo{author}{\bibfnamefont{W.}~\bibnamefont{{Yi}}} \bibnamefont{and}
%  \bibinfo{author}{\bibfnamefont{L.~} \bibnamefont{{Duan}}},
%  \eprint{cond-mat/0605440}.

\bibitem[{\citenamefont{{Norman} et~al.}(1998)\citenamefont{{Norman},
  {Randeria}, {Ding}, and {Campuzano}}}]{norman_pheno_prb_1998}
\bibinfo{author}{\bibfnamefont{M.~R.} \bibnamefont{{Norman} et~al.}},
  \bibinfo{journal}{\prb} \textbf{\bibinfo{volume}{57}}, \bibinfo{pages}{11093}
  (\bibinfo{year}{1998}).

\bibitem[{\citenamefont{Honerkamp et~al.}(2001)\citenamefont{Honerkamp,
  Salmhofer, Furukawa, and Rice}}]{cuprates}
\bibinfo{author}{\bibfnamefont{C.}~\bibnamefont{Honerkamp et~al.}},
  \bibinfo{journal}{Phys. Rev. B} \textbf{\bibinfo{volume}{63}},
  \bibinfo{pages}{035109} (\bibinfo{year}{2001});
\bibinfo{author}{\bibfnamefont{A.~A.} \bibnamefont{{Katanin}}}
  \bibnamefont{and} \bibinfo{author}{\bibfnamefont{A.~P.}
  \bibnamefont{{Kampf}}}, \bibinfo{journal}{Phys. Rev. Lett.}
  \textbf{\bibinfo{volume}{93}}, \bibinfo{pages}{106406}
  (\bibinfo{year}{2004});
\bibinfo{author}{\bibfnamefont{D.}~\bibnamefont{{S{\'e}n{\'e}chal}}}
  \bibnamefont{and} \bibinfo{author}{\bibfnamefont{A.-M.~S.}
  \bibnamefont{{Tremblay}}}, \bibinfo{journal}{Phys. Rev. Lett.}
  \textbf{\bibinfo{volume}{92}}, \bibinfo{pages}{126401}
  (\bibinfo{year}{2004});
\bibinfo{author}{\bibfnamefont{M.}~\bibnamefont{Civelli et~al.}},
  \bibinfo{journal}{Phys. Rev. Lett.} \textbf{\bibinfo{volume}{95}},
  \bibinfo{pages}{106402} (\bibinfo{year}{2005}).

\bibitem{Grimm} C.Chin {\em et al.}, Science {\bf 305}, 1128 (2004).

\bibitem[{\citenamefont{{van Abeelen} et~al.}(1997)\citenamefont{{van Abeelen},
  {Verhaar}, and {Moerdijk}}}]{vanabeelen_li6_pra_1997.pdf}
\bibinfo{author}{\bibfnamefont{F.~A.} \bibnamefont{{van Abeelen} et~al.}},
  \bibinfo{journal}{\pra}
  \textbf{\bibinfo{volume}{55}}, \bibinfo{pages}{4377} (\bibinfo{year}{1997}).

\bibitem{PAD} A. Polkovnikov, E. Altman, E. Demler, PNAS {\bf 103}, 6125 (2006).

\bibitem{NCK} Q. Niu, I. Carusotto, A. B. Kuklov, Phys. Rev. A 73,
053604 (2006).

\end{thebibliography}
\end{document}